\documentclass[pdf,epjc3]{svjour3}  
\smartqed  
\RequirePackage{graphicx}
\RequirePackage{mathptmx}      
\RequirePackage[T1]{fontenc}
\RequirePackage{flushend}
\RequirePackage[numbers,sort&compress]{natbib}
\RequirePackage[colorlinks,citecolor=blue,urlcolor=blue,linkcolor=blue]{hyperref}
\usepackage{amsmath,amssymb}
\usepackage{xcolor}

\usepackage{changes}

\journalname{Eur. Phys. J. C}
\begin{document}

\title{\boldmath Compact stars in $f(T) = T +\xi T^\beta$ gravity}


\author{J. C. N. de Araujo\thanksref{e1,addr1} 
\and 
H. G. M. Fortes\thanksref{e2,addr1,addr2} }                     
%
%

\thankstext{e1}{e-mail: jcarlos.dearaujo@inpe.br}
\thankstext{e2}{e-mail: hemily.gomes@gmail.com}

\institute{Divisão de Astrof\'{i}sica, Instituto Nacional de Pesquisas Espaciais, Avenida dos Astronautas 1758, S\~{a}o Jos\'{e} dos Campos, SP 12227-010, Brazil \label{addr1} \and Instituto Federal Fluminense, Campus Quissam\~{a}, Av. Amilcar Pereira da Silva, 727, Piteiras, Quissam\~{a}, RJ, 28735-000, Brazil.\label{addr2}
}





\date{Received: date / Accepted: date}

\maketitle
\begin{abstract}
{
The Teleparallel Theory is equivalent to General Relativity, but whereas in the latter gravity has to do with curvature, in the former gravity is described by torsion. As is well known, there is in the literature a host of alternative theories of gravity, among them the so called extended theories, in which additional terms are added to the action, such as for example in the $f(R)$ and $f(T)$ gravities, where $R$ is the Ricci scalar and $T$ is the scalar torsion, respectively. One of the ways to probe alternative gravity is via compact objects. In fact, there is in the literature a series of papers on compact objects in $f(R)$ and $f(T)$ gravity. In particular, there are several papers that consider $f(T) = T + \xi T^2$, where $\xi$ is a real constant. In this paper, we generalise such extension considering compact stars in $f (T ) = T + \xi T^\beta$ gravity, where $\xi$ and $\beta$ are real constants and looking out for the implications in their maximum masses and compactness in comparison to the General Relativity. {Also, we are led to constrain the $\beta$ parameter to positive integers which is a restriction not imposed by cosmology.}}

\end{abstract}

\section{Introduction}
\label{int}

As is well known,  General Relativity (GR) is a well-tested theory of gravity over a large range of different energies and its predictions have been confirmed even today. {Nevertheless, there are
some theoretical impasses such as the cosmological constant problem \cite{Weinberg} and
the unexplained accelerated expansion of the Universe \cite{Brax}, which have urged the search for alternative models that could overcome them, describing gravitation on cosmological scales and, in addition, reproducing in certain limits the results of GR \cite{intro1,intro2,intro3,intro4,intro5,intro6}}.

{In order to reproduce the known results and go beyond, one of the simplest modifications to GR would be the so-called $f(R)$ models \cite{fR} where we have a function of the scalar of curvature $R$ as Lagrangian density.} {Many applications of $f(R)$ were considered recently in different scenarios, whether presenting cosmological solutions or modelling astrophysical objects such as compact stars. See \cite{fR3,fR4,fR5,fR6,fR7,fR8,fR9,fR10,fR11,fR12,fR13} for some recent references.}

{Additionally, one can have alternative descriptions of gravity, such as the named Teleparallel theory of Gravity \cite{Aldro,Bahamonde,TT} which is equivalent to GR, with the basic difference of having in its formulation the torsion scalar $T$ instead of the curvature scalar R. {The torsion is formed from products of first derivatives of the tetrad, with no second derivatives appearing in the torsion tensor.} In the same way as the GR can be extended through the $f(R)$ models {which has attracted much attention in recent years as a way to explain acceleration of the universe \cite{fR2}}, the natural extension for the Teleparallel theory of gravity is the $f(T)$ models \cite{Ferraro,Linder} \textcolor{black}{with the advantage of keeping its field equations second order, unlike the fourth order equations of $f(R)$ theory.} The study of these models in which the gravity is described by the torsion $T$ instead the curvature $R$ is growing more and more. There is a host of papers which consider, for example, $f(T)$ gravity on cosmology and compact object models \cite{Ganiou,Kpa,Pace,Ilijic,Bohmer,Pace2,FA1,FA2,FA3,FA4,Myr,Karami,SSS10,Ilijic2}}.

{One of the possible consequences in considering different gravity $f(T)$ models is to determine the maximum mass allowed for the star for a given equation of state which could, in principle, lead us to different maximum masses, see \cite{Nunes}. This would be important, for example, to explain the secondary components in the event GW190814 \cite{GW} as a candidate to a relatively high-mass neutron star. As is well known, in GR there is the so-called Buchdahl limit \cite{Dadhich,buch1959} for the compactness of a compact star where $M/R \leq 4/9$. Thus, a natural question to ask would be whether a less restrictive compactness can be achieved from an alternative theoretical description.}

{We have been dealing with this issue lately and gathering some results in a series of papers \cite{FA1,FA2,FA3,FA4}, where a novel approach to obtain the Generalized Tolman-Oppenheimer-Volkoff (TOV) equations \cite{TOV} {is presented, } which are derived from {a particular $f(T)$} gravity and can be used to model the structure of a spherically symmetrical object that is in hydrostatic equilibrium. Then, these generalized TOV equations are used for modelling neutron stars and applied to polytropic and realistic equations of state.}

\textcolor{black}{Once gravity is modified, a fundamental issue arises. In alternative descriptions of gravity, it is not guaranteed that the mass of a compact object can be calculated in the same way as in GR. We have properly discussed this problem of mass definition in theories with torsion in \cite{FA4}. In summary, whereas the mass of compact objects in GR is unambiguously given, this is not necessarily the case in some alternative theories of gravity \cite{olmo}. The Schwarzschild metric is not assumed as a exterior solution in our approach. So, it was necessary to investigate different ways to calculate mass of compact stars in $f(T)$ gravity and identify among them the most appropriated, which seems to be the mass measured by an observer at infinity.}

In all those works, the model taken into account was $f(T)=T+\xi \, T^2$ where $\xi$ is an arbitrary real constant \cite{Ilijic,Ilijic2,Ilijic20,Nair,FA2,FA3}. The results have validated the idea that greater maximum masses and less restrictive compactness can be achieved from alternative models. \textcolor{black}{The most common extension of the teleparallel gravity in the literature is precisely the addition of a quadratic term, either for the simplicity or for the inspiration of the Starobinsky model in $f(R)$. In this sense, it is natural to perform the same analysis for more general models such as $f(T)=T+\xi \, T^\beta$ where now $\beta$ is also an arbitrary real constant, which is our main purpose here. There are several works in the literature which considers a power-law approach for $f(T)$ extension in different scenarios, such as, the study of the detailed dynamics of the cosmological evolution \cite{TT}, the accelerated expansion without dark energy but driven by torsion \cite{Bengo} and also the power-law model can leads to an asymptotic future de Sitter state for the universe in the study of the cosmic acceleration \cite{Linder}. Thus, some studies in cosmology already have taken into account this functional form with $\beta \neq 2$. Therefore, it does make sense to extend this analysis for modelling compact objects which will lead us to constrain this parameter, as will be seen further.}



\textcolor{black}{In Section \ref{be}, we present the main equations for modelling spherically symmetric distributions from $f(T)$ gravity, more specifically, $f(T)=T+\xi \, T^\beta$. Also, the set of differential equations derived from the extended action is presented. In Section \ref{models}, we explain how is the guideline to perform the analysis for a compact star using the equations from previous section and how it can be done using Python programming. The issue regarding the most appropriate mass definition for $f(T)$ gravity is also discussed. In Section \ref{nr}, the numerical results for a given polytropic equation of state by considering different values for $\xi$ and $\beta$ and also sequences of models representing the mass $M$ as a function of the radius $R$ and central energy density $\rho_c$ are present and their implication on the $\xi$ and $\beta$ parameters are discussed. Finally, in Section \ref{conclusions}, the main results are summed up.}


\section{The basic equations of f(T) gravity for spherically symmetric metric}
\label{be}

Since our interest here is modelling spherical stars where rotation is not taken into account, let us consider the spherically symmetric metric \cite{SSS11,olmo,Ilijic20,SSS12,Bohmer}:
\begin{equation}
    ds^2=e^{A(r)}\, dt^2-e^{B(r)}\, dr^2-r^2\, d\theta^2-r^2 \sin ^2 \theta \, d\phi^2. \label{metric}
\end{equation}
As is well known, the dynamical variable for torsion models is the tetrad ${h^a}_\mu$ which is related to the metric via $g_{\mu\nu}={h^a}_\mu {h^b}_\nu \, \eta_{ab}$, where $\eta_{ab}$ is the mostly minus Minkowski metric. The choice of the tetrad can be a delicate issue which we have discussed in \cite{FA1}. For the calculations here, we are going to adopt the following tetrad \cite{Tamanini}:
\begin{eqnarray}
\setlength{\arraycolsep}{3mm}
{h_\mu}^a=
\left(
\begin{array}{cccc}
 e^{A/2} & 0 & 0 & 0\\
 0 & e^{B/2} \sin \theta \cos \phi  & e^{B/2} \sin \theta \sin \phi & e^{B/2} \cos \theta \\
 0 & r \cos \theta \cos \phi & r \cos \theta \sin \phi & -r \sin \theta\\ 
 0 & - r \sin \theta \sin \phi & r\sin \theta \cos \phi & 0
\end{array}\right).
\label{006}
\end{eqnarray}
{
It is worth noting that although we are not using the covariant formulation \cite{Krssak} here, Ref. \cite{Tamanini} has showed that the choice of nondiagonal (rotated) tetrads, {like the one above}, permit us to recover spherically symmetric solutions in vacuum, using the equations presented here derived from the original formulation. {In short, the tetrad above generates the same set of equations in both the original and covariant formulations.}}

{As mentioned in the introduction, the $f(T)$ model to be considered in this work is the following:
\begin{eqnarray}
f(T)=T + \xi \, T^\beta \, ,
\label{fTbeta}
\end{eqnarray}
where $\xi$ and $\beta$ are arbitrary real numbers.}

{The expression for the scalar torsion $T$ is well known and can be obtained from the antissymmetric part of the connection, which leads to:
\begin{equation}
  T(r) = \frac{2\, e^{-B}}{r^2} 
  \Bigl[ 
    1 + e^B - 2\,e^{B/2}  + 
        r\, A' \left(1-e^{B/2}\right) 
  \Bigr]\,.
  \label{008}
\end{equation}}

{The equations of motion can be obtained from the variation of the action for the $f(T)$ model
\begin{eqnarray}
    S=\int \biggl(\frac{f(T)}{16\pi}+\mathcal{L}_m\biggr) \, \mbox{det}({h_\mu}^a) \,  d^4 x
\end{eqnarray}
with respect to the Vierbein components ${h_\mu}^a(r)$, where $\mathcal{L}_m$ is the Lagrangian density of matter fields. By doing so, one obtains a non-trivial system of equations involving $A'$, $B$, $B'$, $\xi$, $\beta$, $P$ and $\rho$. The pressure $P$ and mass density $\rho$ arise due the momentum-energy tensor and an equation of state (EOS) relates $P$ and $\rho$. For more details, see \cite{FA1} where we have considered the case $\beta=2$.}

{From the above action, we then obtain a set of equations, whose manipulation leads us to the following set of equations to be numerically solved: }
\begin{eqnarray}
P =&\,& \frac{{\rm e}^{-B}}{16\pi r^2}\left[ 2\left(1-{{\rm e}^{B}}  + rA'\right)\left(1 + \xi\beta T^{\beta -1} \right)+ (\beta-1)r^2{\rm e}^B \xi T^\beta  \right],
\label{press}
\\
&\,& \nonumber
\\
B' = &\,&  \frac{{{\rm e}^{-B}}}{2r^3}\bigg\{1 + \beta\xi T^{\beta -1}\left[2\beta + (2\beta-1)\beta\xi T^{\beta -1}\right]\bigg\}^{-1}\nonumber \\
&\,& \cdot \ \bigg\{2r^2{\rm e}^{2B}\left[8\pi\rho r^2\left(1+\xi\beta T^{\beta -1}\right)+{\rm e}^{-B} -1 \right]
\nonumber \\
&\,&+\ 32\pi\rho r^2\xi\beta(1-\beta)T^{\beta-2}{\rm e}^{B}({\rm e}^{1/2\,B}-1)(rA'-2{\rm e}^{1/2\,B}+2)
\nonumber \\
&\,&+ \ r^4\xi(\beta-1)T^\beta{\rm e}^{2B}\left(1+\xi\beta T^{\beta -1}\right)\nonumber \\
&\,& + \ 4\xi\beta(1-\beta)\left(1 + \xi\beta  \right)T^{2\beta -3} ({\rm e}^{B/2}-1)^2 \left[6({\rm e}^{B}-1) -rA'({\rm e}^{B/2}+5)+
r^2A'^2 \right]
\nonumber \\
&\,&+ \ 4r^2\xi\beta T^{\beta-1}{\rm e}^{B}({\rm e}^{B/2}-1)\left(2\beta-3-{\rm e}^{B/2}\right)
\nonumber \\
&\,& +\  2 \beta (\xi r) ^2{\rm e}^{B}({\rm e}^{B/2}-1)T^{2\beta-2} \cdot \nonumber \\
&\,& \cdot \ \left[ 2({\rm e}^{B/2}-1)-rA'(\beta-1)^2+2\beta^2({\rm e}^{B/2}+1)-\beta(5{\rm e}^{B/2}+1) \right]\bigg\} \ , \label{B'}\\
&\,& \nonumber
\\
A' = &\,& -2 \frac{P'}{P+\rho},
\label{ce1}    
\end{eqnarray}
where (\ref{ce1}) is the well-known {\it conservation equation} \cite{Bohmer}.

\section{Compact stars on $f(T)$ }
\label{models}

{To model spherical stars on $f(T)$ one can follow, e.g., the prescription provided by Ref. \cite{FA1}, with one exception: in \cite{FA1}, in which $f(T) = T + \xi T^2$, $A'$ is obtained with an expression without needing to perform  a numerical integration of a differential equation. Here, it is not necessary to integrate a differential equation for $A'$ as well, but one needs to solve numerically a system formed by the two nonlinear equations (\ref{008}) and (\ref{press}) to obtain it.  For given $\xi$, $\beta$, $P$, $B$
and $r$, the numerical solution of the referred system provides $A'$ and $T$ at $r$.} \textcolor{black}{It is worth noting that the system does not depend on $A$, but only on $A'$, which offers some advantage in the numerical integration of the system of equations above.}


{We have written a numerical code in Python to solve the system of equations shown in previous section for a given EOS. To do so, it is necessary to provide the following boundary conditions
\begin{equation}
     P = P_c \quad {\rm at} \quad r =0,
\end{equation}
from which one obtains $P(r)$ and $\rho(r)$, i.e., the structure of the star. The radius of the star, $R$, is the value of $r$ for which $P(r) = 0$. In short, one starts the integration at $r=0$ and continues it till the value of $r$ for which $P(r)=0$ at the star's surface.}

{Here, unlike General Relativity, there is a differential equation for $B(r)$. Thus, a central boundary condition for $B$ must be given. In this sense, regularity at $r=0$ implies that $B_c = 0$.} \textcolor{black}{Since our system does not depend explicitly on $A$, it is not necessary to set a condition for $A(r=0)$.}


{
In General Relativity, to model compact stars such as neutron stars, the equation commonly considered for the calculation of mass interior to radius $r$, namely $m(r)$, is
\begin{eqnarray}
    \frac{dm}{dr}=4\pi \rho r^2 \ .
    \label{MGR}
\end{eqnarray}
However, when we are dealing with alternative description of gravity, one must address this issue carefully.

\textcolor{black}{The problem is that unlike the GR where the mass is unambiguously given, in $f(T)$ models this is considered an open question. The internal energy and the gravitational potential energy enter in the calculation of mass in (\ref{MGR}). Since gravity is now modified, there is no guarantee that (\ref{MGR}) accounts for the mass of a compact object in $f(T)$ gravity. This can be seen from equation for $B'$ in (\ref{B'}) which holds inside and outside the star, since $\rho$ goes to zero smoothly. Then, the behaviour of $B$ outside the star, for example, is affected by the particular $f(T)$ considered. We do not impose any functional form for $B(r)$, as some authors do. This function is always obtained via the numerical integration of equation for $B'$ in our calculations. Consequently, the Schwarzschild metric is not assumed as a exterior solution in our approach.}
\textcolor{black}{So, it is necessary to consider different and consistent ways to measure the mass in $f(T)$ and we have discussed this question in details in \cite{FA4}. In particular, we argue that the asymptotic mass may be the most appropriate way to calculate mass in this theory.}

In summary, while the mass of compact objects in GR is a well-defined quantity, in $f(T)$ there may be ambiguity in its definition \cite{olmo}. However, the most appropriated way to calculate mass in $f(T)$ gravity seems to consider the mass measured by an observer at infinity:
\begin{equation}
    M_{\infty} =  \lim_{r\to\infty} \frac{1}{2}\, r\,B(r)\ .
\end{equation}

\textcolor{black}{Also, as is well known, in the Teleparallel theory of Gravity, the Ricci scalar outside (vacuum) of any matter distribution is zero. One would think of that the torsion would be also null. However, having a look, for example, at equation (\ref{008}), one can easily conclude that $T$ is not null outside a spherically symmetrical distribution which affects the calculation of mass. Thus, an observer near to the star feels the contribution of $M_\infty$ and additional terms that come from the $T^\beta$ term. A deeper discussion regarding the case of $f(T)=T+\xi T^2$ can be found in \cite{FA1}}.

{Another interesting quantity to be calculated, although not an observable, is the total rest mass $M_0$, namely
\begin{eqnarray}
\frac{dM_0}{dr}=4\pi \rho_0 \,e^{B/2}\, r^2 \ ,
\label{dm0dr}
\end{eqnarray}
where $\rho_0$ is the rest mass density. Starting from a given EOS
and a central density, different theories lead to different $M_0$ which allows us to compare them.}


\textcolor{black}{All this is due to the fact that, in $f(T)$ gravity, the space-time exterior to a non-rotating and spherically symmetric object is not available in closed form as in GR. Thus, the vacuum space-time is not given by the Schwarzschild metric. In the literature, we can find only a perturbative expression in the weak gravity regime and also an approach with non-Schwarzschild vacuum.}

\begin{figure}
    \centering
    \includegraphics[width=0.48\linewidth]{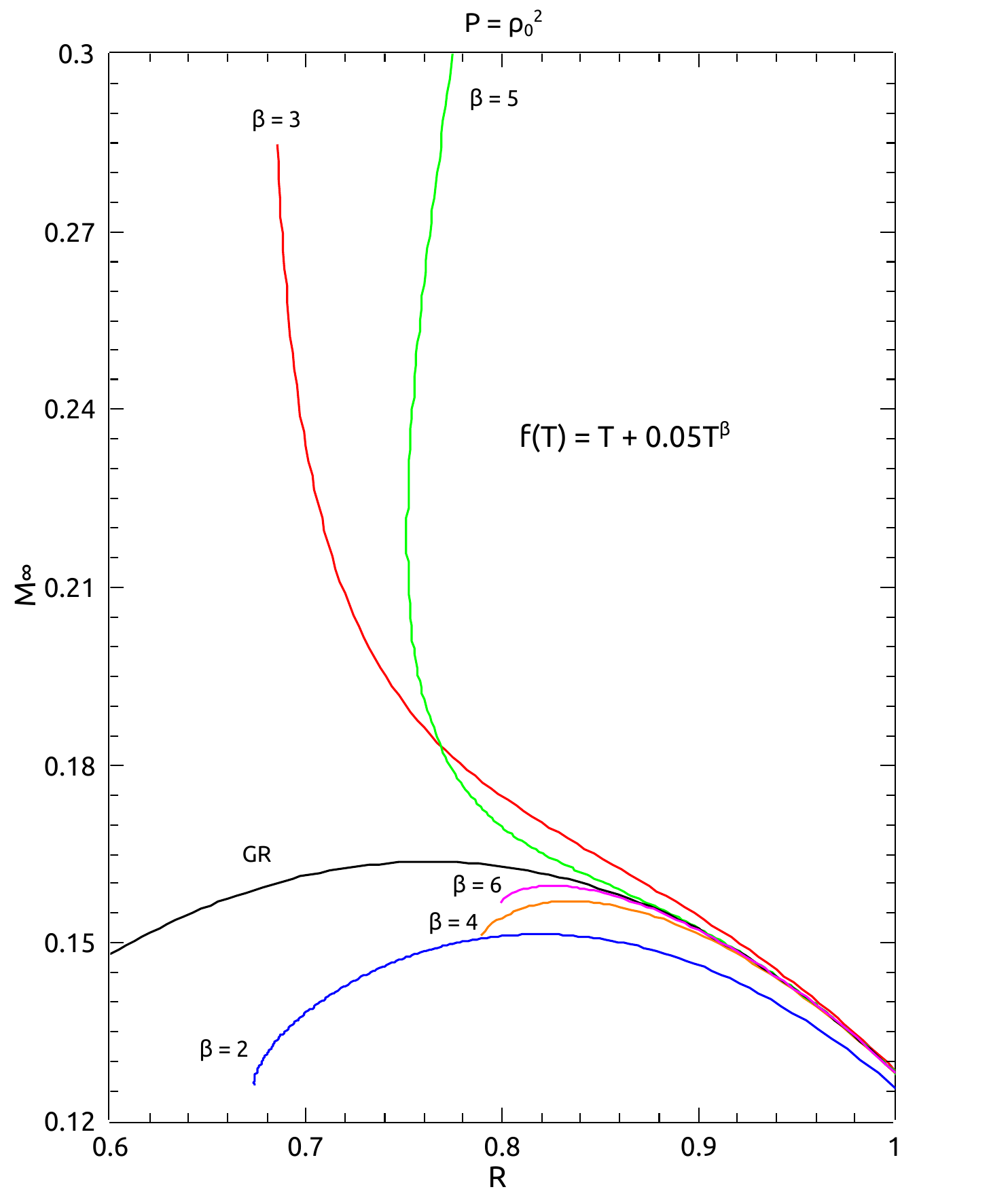}
    \centering
    \includegraphics[width=0.48\linewidth]{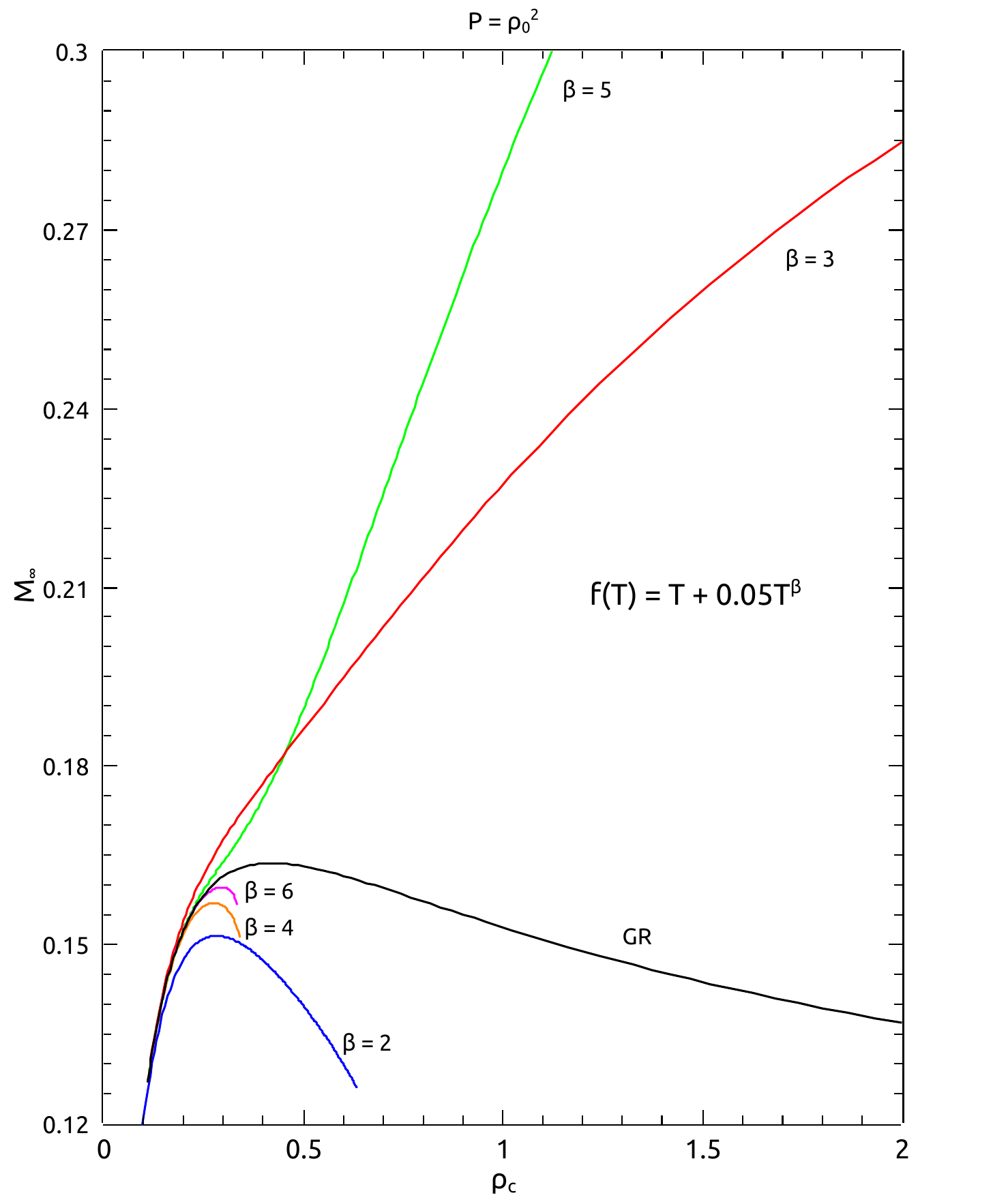}
    \caption{Sequences of mass M as a function of the radius R (left panel) and the central energy density $\rho_c$ (right panel) for $\xi = 0.05$ and different values of $\beta$.}
    \label{Betas1}
\end{figure}

\begin{figure}
    \centering
    \includegraphics[width=0.48\linewidth]{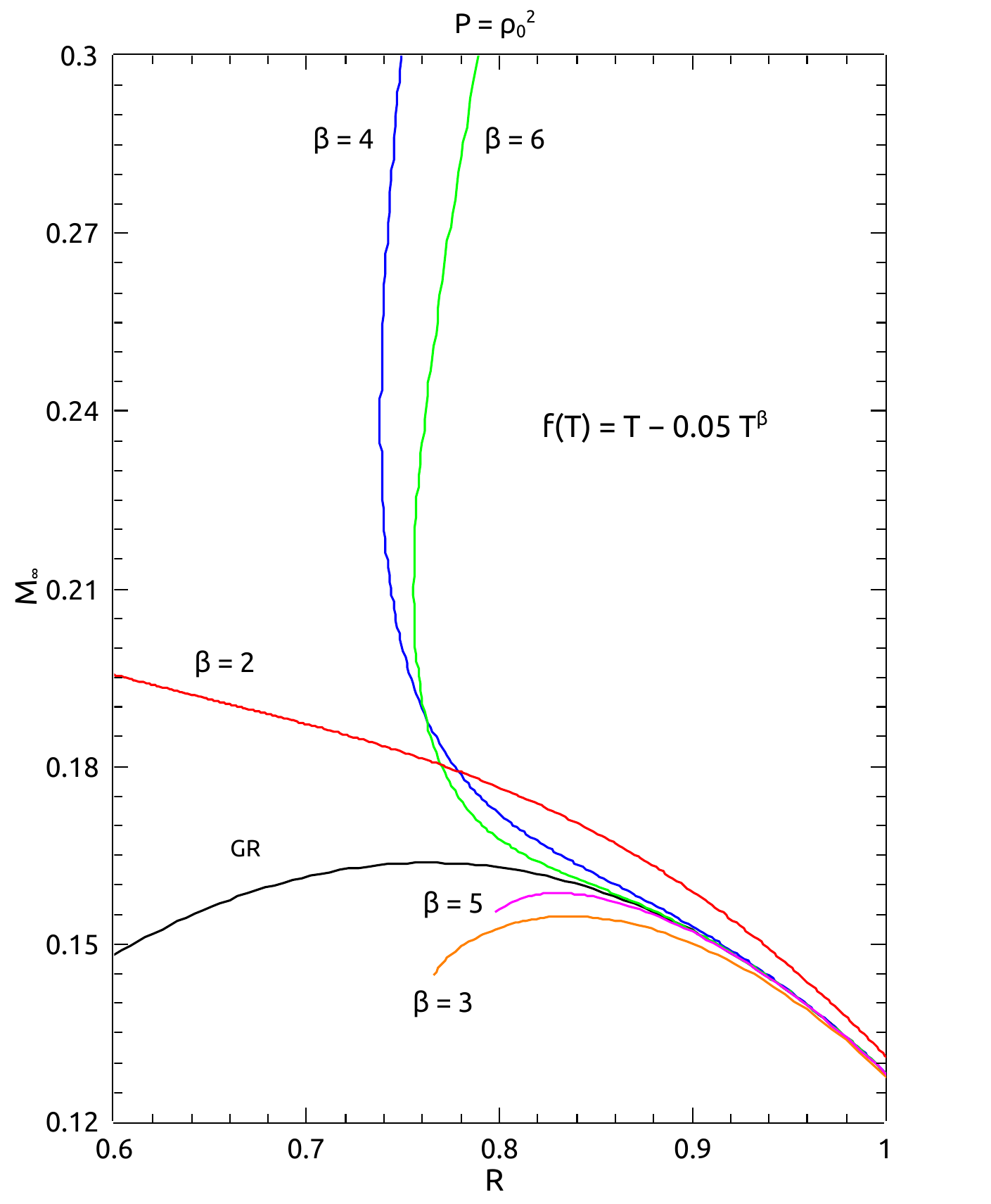}
    \centering
    \includegraphics[width=0.48\linewidth]{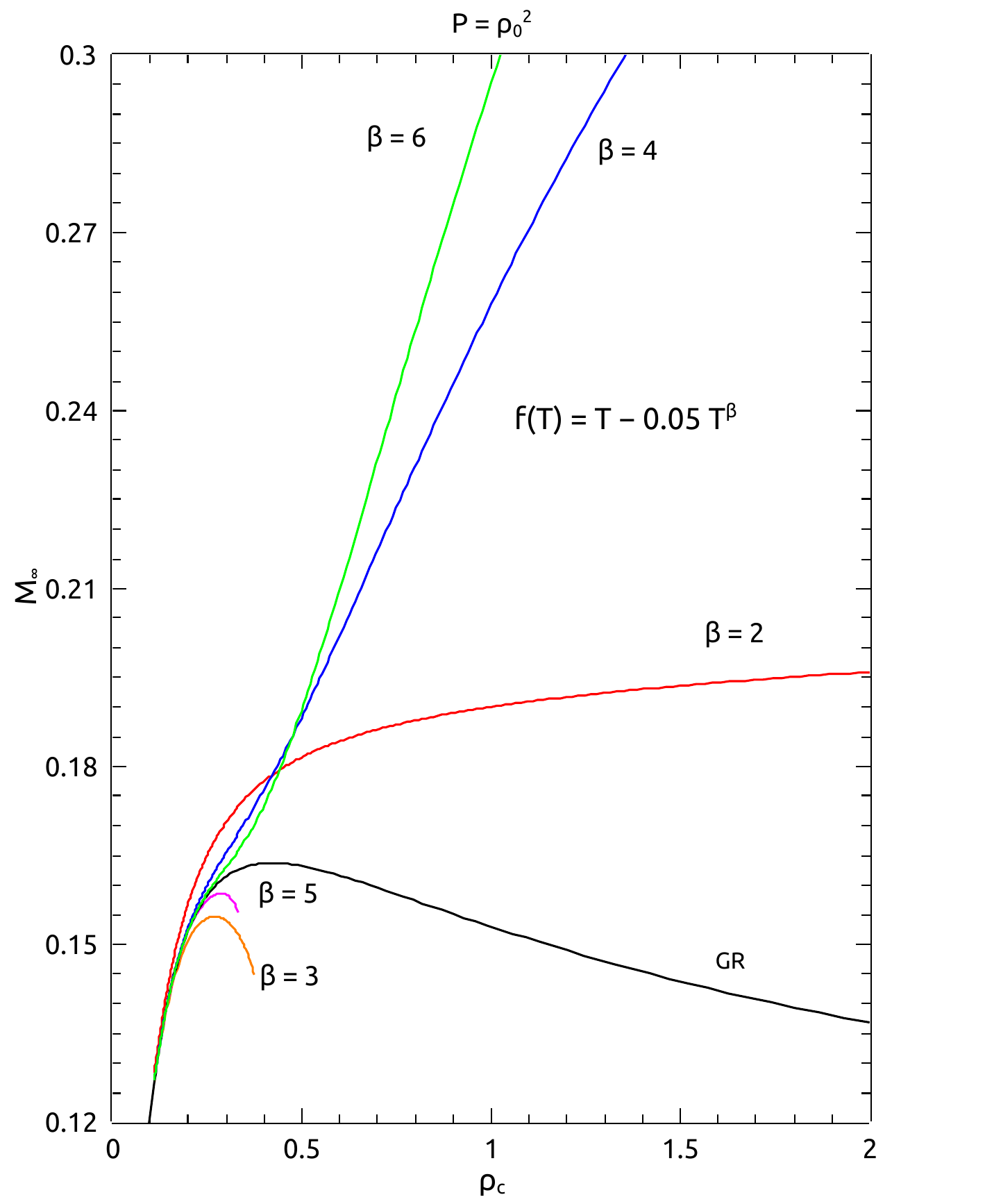}
    \caption{The same as in figure \ref{Betas1} now for $\xi = -0.05$.}
    \label{Betas2}
\end{figure}

\begin{figure}
        \centering
        \includegraphics[width=0.48\linewidth]{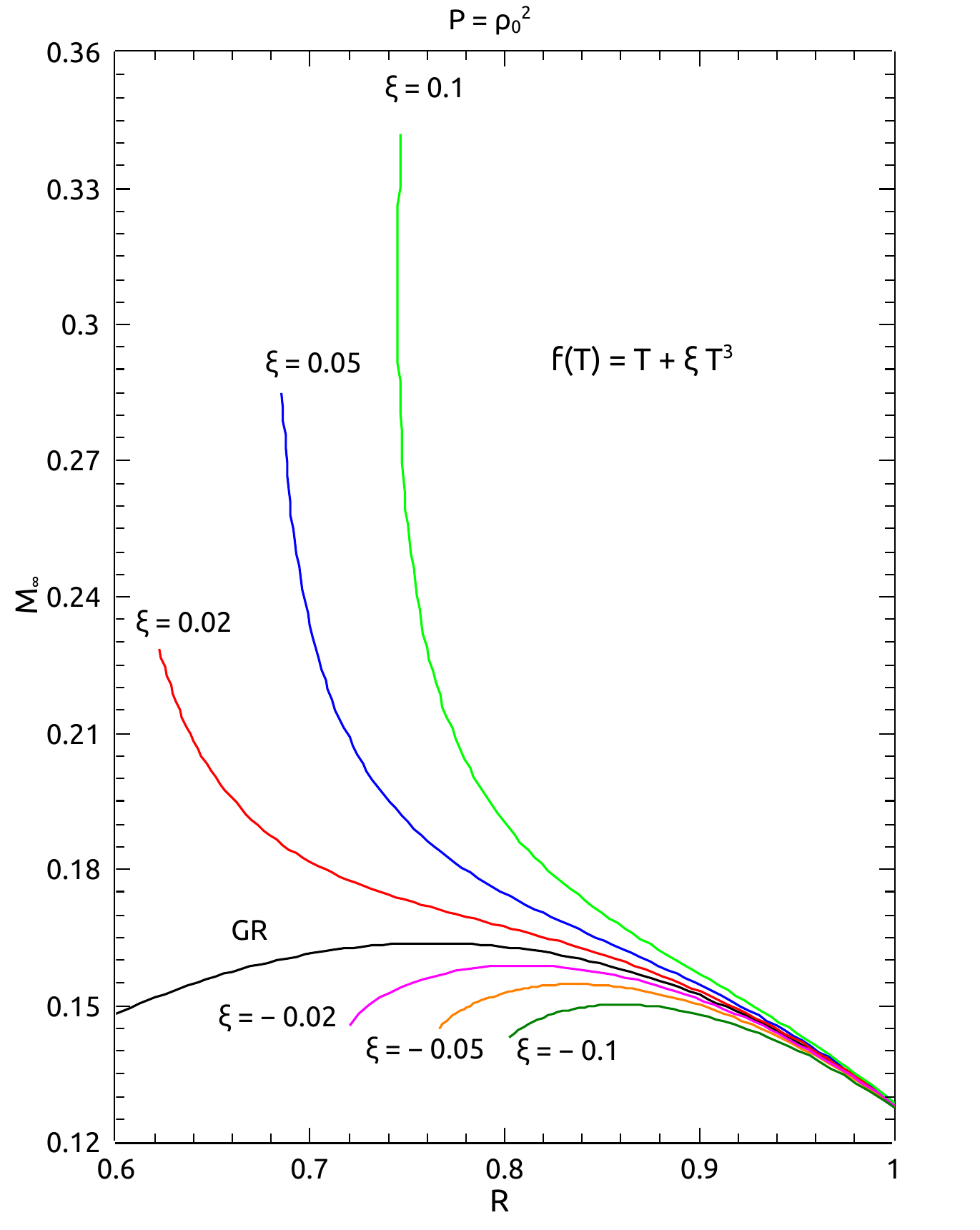}
        \centering
        \includegraphics[width=0.48\linewidth]{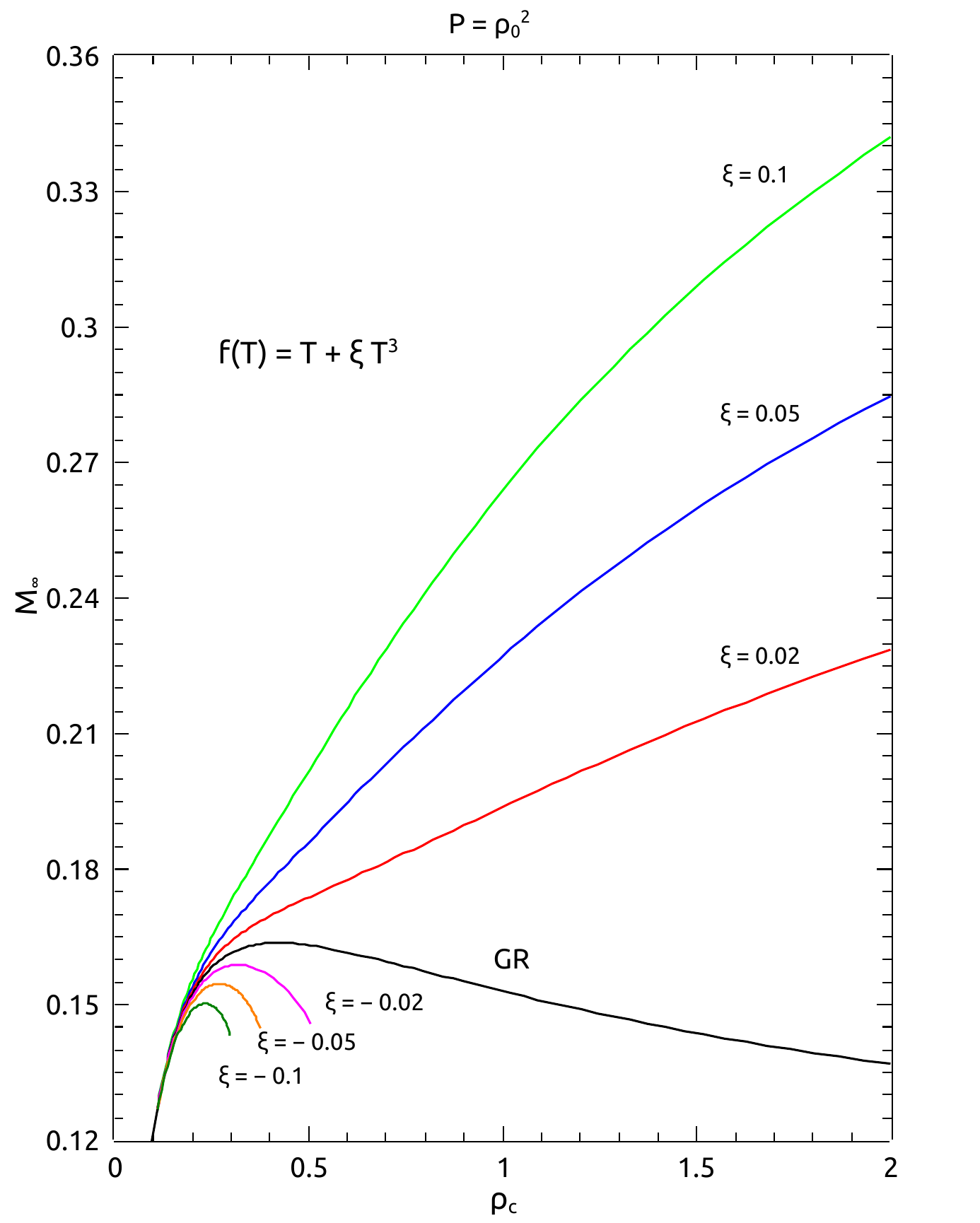}
        \caption{Sequences of mass M as a function of the radius R (left panel) and the central energy density $\rho_c$ (right panel) for $\beta = 3$ and different values of $\xi$.}
    \label{Beta3}
\end{figure}

\begin{figure}
    \centering
    \includegraphics[width=0.48\linewidth]{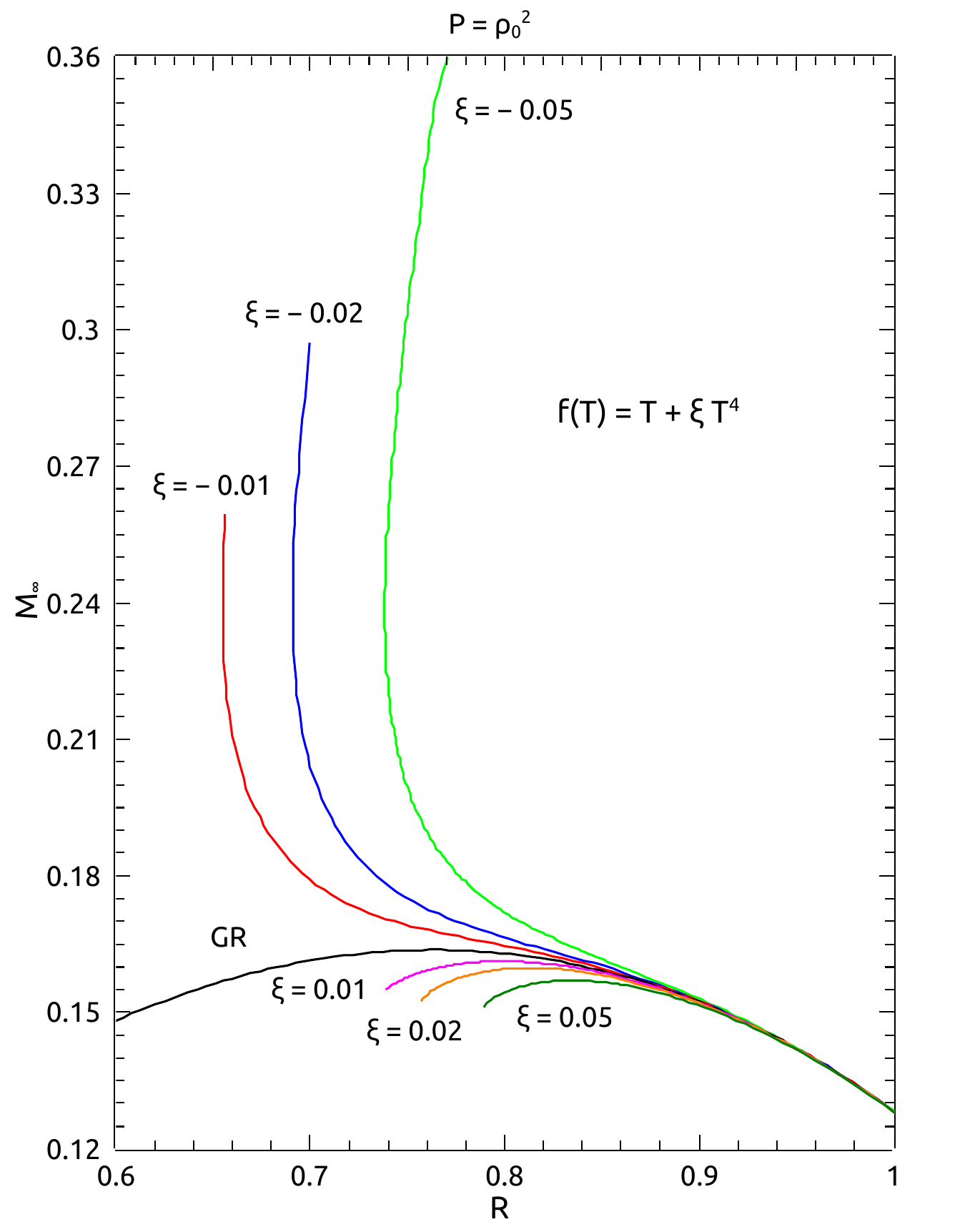}
    \centering
    \includegraphics[width=0.48\linewidth]{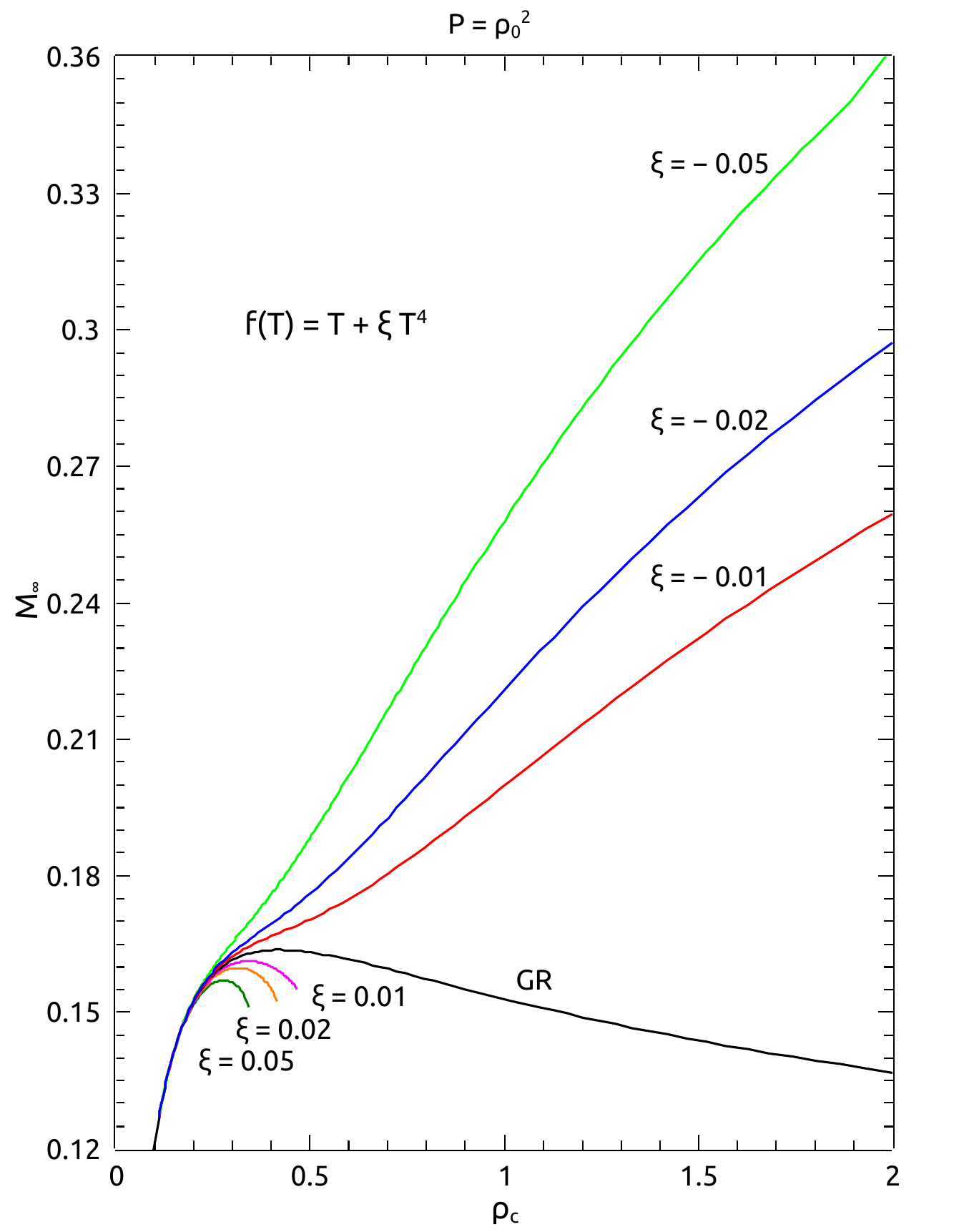}
    \caption{The same as in figure \ref{Beta3} now for $\beta =4$.}
    \label{Beta4}
\end{figure}

\section{Numerical results} \label{nr}

\textcolor{black}{In this section, some numerical calculations for different values of $\xi$ and $\beta$ are shown. For this purpose, we first choose an equation of state and, using the appropriate numerical integrator in Python, we solve the set of differential equations seeking for the values of $M_\infty$ and $R$, given a central energy density $\rho_c$. From that, we can also build the so-called sequences of models representing the mass $M_\infty$ as a function of the radius $R$ and central energy density $\rho_c$ and compare them.} \textcolor{black}{To begin  with, we adopt a polytropic EOS \cite{eos0}, which is very useful in comparing models with different values of $\xi$ and $\beta$.}

{Recall that the polytropic is given by
\begin{equation}
    P = k\, {\rho_0}^ \gamma,
    \label{tpol}
\end{equation}
where $\rho_0$ is the rest-mass density, $k$ is the polytropic gas constant and $\gamma$ is the adiabatic index, which is related to the polytropic index $n$ via $\gamma \equiv 1+1/n$. The mass-energy density, $\rho$, namely, $\rho = \rho_0 + n\,P$ is obtained (see, e.g., \cite{BS2010}) from the first law of thermodynamics. In particular, we consider $n =1$, which corresponds to $\gamma = 2$, in all models studied here. As in Refs. \cite{FA1,FA2},  we set all quantities in dimensionless form.\footnote{{In \cite{BS2010} it is shown that, in geometrized units, $k^{n/2}$ has units of length. Therefore, one can define nondimensional quantities, namely, $\bar{r} = k^{-n/2}r $, $\bar{P} = k^{n}P $, $\bar{\rho} = k^{n}\rho $, $\bar{M} = k^{-n/2}M $ and $\bar{T} = k^{n}T $. To simplify the notation, we then omit the bars in our equations. }} In practice, this is equivalent to set $k = G = c = 1$. 
}

{
Before proceeding, it is worth considering what values the $\beta$ parameter can have. First, $\beta$ cannot be negative, otherwise $f(T)$ could have extremely large values and even diverge. Second, since the torsion is negative (see, e.g., Ref. \cite{FA1}) and observing the way the torsion appears in equations (\ref{press}) and (\ref{B'}), one concludes that $\beta$ must be a positive integer. Then, we shall consider in particular integer values of $\beta$ ranging from two to six, which give us a good idea of how different values of $\beta$ modify the $M$ vs $R$ sequences. It is worth noting that, regarding the parameter $\xi$, there is no restriction as long as it is a real number.
}

{
In figure \ref{Betas1}, ``Mass $\times$ Radius'' and ``Mass $\times$ $\rho_c$'' sequences for $\xi = 0.05$ and integer values of $\beta$ in the range $[2,6]$ are shown, where $\rho_c$ is the central density.
}

{
An obvious fact is that even and odd values of $\beta$ have very different behaviours. Whereas even  $\beta$'s  generate sequences bellow the GR sequence and present maximum masses, odd $\beta$'s
generate sequences above that of GR and do not have maximum masses.
}

{
As discussed, for example, in Ref. \cite{FA1}, the torsion is related to the intensity of the gravitational interaction. The more negative the torsion, the more intense the gravitational interaction. Thus, for $\xi > 0$ the torsion is more negative for odd $\beta$'s. As a result, the gravitational interaction is more intense. Consequently, the objects can be more massive. 
}

{
Also note that the objects modelled with the odd $\beta$'s studied here can be much more compact than those predicted by GR.} {This can be seen by looking at the curves above that from GR in Figure \ref{Betas1}. Within the same value of $R$, for example, $R=0.8$, there is more mass when we have odd $\beta$'s, i.e., more compactness.}

{
The models for crescent values of even $\beta$'s also present interesting results.
For example, the sequences are closer to that predicted by GR, although with lower maximum masses and central densities $\rho_c$.
}

{
In figure \ref{Betas2} we have the same as in figure \ref{Betas1} now for $\xi = -0.05$. In general, the conclusions are similar to those for $\xi = 0.05$, but the roles of odd and even $\beta$'s are reversed.
}

{
Now we consider a set of sequences for $\beta = 3$ and 4 for different values of $\xi$.  First, in figure \ref{Beta3}, sequences for $\beta = 3$ and different values of $\xi$ are shown. Note that the sequences are similar in form to those of the well studied case of $\beta = 2$, although now the objects can well be much more compact.
}

{
Finally, in figure \ref{Beta4}, sequences for 
$\beta =4$ are shown. Our calculations show that the objects are even more compact than predicted for $\beta = 3$. Moreover, the calculations are even more sensitive to $\xi$.}

{A brief qualitative summary about the behaviour of the models for different $\beta$ and $\xi$ are displayed in Table \ref{tabela}.}

\begin{table}[h!]
\centering
\renewcommand{\tabcolsep}{15 pt}
\renewcommand{\arraystretch}{1.5}
\begin{tabular}{|c|l l|}
\hline
$\beta$ &   even (odd) & odd (even) \\
\hline
& sequences below GR & sequences above GR\\
$\xi>0$ & maximum mass & no maximum mass \\
($\xi <0$) & less negative torsion & more negative torsion \\
 & less compactness & more compactness \\
\hline
\end{tabular}
\caption{Some features of the models according to the $\beta$ and $\xi$ values}
\label{tabela}
\end{table}

\section{Final remarks}
\label{conclusions}


{
As is well known,  if $f(T)=T$, one has the Teleparallel theory which is completely equivalent  to General Relativity. Additional terms such as $\xi \, T^\beta$ provide a simple way to obtain an extended  Teleparallel theory, {as well as what happens with $f(R)$ for extended curvature models.}
}

There is in the literature a series of papers modelling spherical stars in $f(T)$ gravity \cite{Ganiou,Kpa,Pace,Ilijic,Bohmer,Pace2,SSS11}. Notice, however, that many authors consider a very restrict functional form for $f(T)$, namely, $f(T) = T + \xi T^2$, \textcolor{black}{which is indeed the most natural first extension of the Teleparallel gravity for its simplicity and motivated by the Starobinsky model in $f(R)$. A relevant question is what would be the influence of the power of $T$ and, therefore, consider different functional forms for $f(T)$ in order to study how the results and conclusions depend on the particular form adopted.} \textcolor{black}{Here, we take into account such issue,} considering $f(T) = T + \xi T^\beta$, where $\xi$ and $\beta$ are real constants.

\textcolor{black}{As already discussed, due to the fact that in $f(T)$ gravity the spacetime exterior to a non-rotating and spherically symmetric object is not given by Schwarzschild metric, the mass definition is ambiguous. So, for all the calculations performed here, we have considered the mass measured by an observer at inifinity, $M_\infty$ \cite{FA4}.}


\textcolor{black}{Regarding to the numerical results,} we argue that $\beta$ must be an integer constant in order to be possible to solve the set of differential equations used to model spherical stars. On the other hand, $\xi$ can assume any real number. \textcolor{black}{Notice that we are led to constrain the $\beta$ parameter, which is a restriction that does not occur in cosmology.}

{
Considering models for $\xi > 0$ ($\xi < 0$) and crescent odd (even) values for $\beta$, which provide $\xi T^\beta < 0$, the resulting stars are increasingly compact, as can be inferred from figures \ref{Betas1} and  \ref{Betas2}. Also, for the values of $\xi$ studied, there is no maximum mass if $\xi T^\beta < 0$.}

\textcolor{black}{On the other hand,} models for $\xi T^\beta > 0$ generate sequences bellow those of General Relativity. These sequences, unlike those for $\xi T^\beta < 0$, present maximum masses, although lower than that of General Relativity.

\textcolor{black}{Those results show that, indeed, the functional form of $f(T)$ in power-law of $T$ can change significantly the limits of mass and compactness of the star.} {An interesting next step is to consider, for example, $f(T) = T + \alpha T^2 + \sigma T^3$ or other combinations of power laws which may perhaps bring up some different behaviour for the modelling of compact objects. We shall consider such an issue in another paper to appear elsewhere.}

\begin{acknowledgements}
J.C.N.A. thanks CNPq (307803/2022-8) for partial financial support. H.G.M.F. thanks CNPq for the financial support (152326/2022-7).
The authors would like to thank Rafael da Costa Nunes for discussions related to the $f(T)$ theory.
\end{acknowledgements}



\begin{thebibliography}{99}

\bibitem{Weinberg} S. Weinberg. \emph{The cosmological constant problem}. Reviews of Modern Physics 61(1), pp.1-23 (1989).

\bibitem{Brax} P. Brax. \emph{What makes the Universe accelerate? A review on what dark energy could be and how to test it}. Reports on Progress in Physics, 81(1), p.016902 (2017).

\bibitem{intro1} T. Clifton, P. G. Ferreira, A. Padilla, C. Skordis. \emph{Modified Gravity and Cosmology}, Physics Reports 513 (2012) 1 [arXiv:1106.2476].


\bibitem{intro2} S. Nojiri, S.D. Odintsov. \emph{Unified cosmic history in modified gravity: from F(R) theory to Lorentz non-invariant models}. Phys. Rept. 505 (2011) 59 [arXiv:1011.0544].

\bibitem{intro3} 

S. Capozziello, M. De Laurentis. \emph{Extended Theories of Gravity}. Physics Reports, v. 509, n. 4, p. 167–321, (2011) [arXiv:1108.6266].
‌

\bibitem{intro4} S. Nojiri, S. D. Odintsov, V. K. Oikonomou. \emph{Modified Gravity Theories on a Nutshell: Inflation, Bounce and Late-time Evolution}.  Physics Reports, v. 692, p. 1–104, (2017) [arXiv:1705.11098].

\bibitem{intro5} T. P. Sotiriou, V. Faraoni. \emph{f(R) Theories Of Gravity}. Rev. Mod. Phys. 82 (2010) 451 [arXiv:0805.1726].

\bibitem{intro6} S. D. Odintsov, V. K. Oikonomou, I. Giannakoudi, F. P. Fronimos, E. C. Lymperiadou. \emph{Recent Advances on Inflation}. Symmetry 15 (2023) 1701 [arXiv:2307.16308]

\bibitem{fR} T. P. Sotiriou and V. Faraoni. \emph{$f(R)$ theories of gravity.} Rev. Mod. Phys. 82, 451 (2010).

\bibitem{fR3} A. de la Cruz-Dombriz,  D. Saez-Gomez. \emph{Black Holes, Cosmological Solutions, Future Singularities, and Their Thermodynamical Properties in Modified Gravity Theories}.  Entropy, v. 14, n. 9, p. 1717–1770, (2012) arXiv:1207.2663 [gr-qc]

\bibitem{fR4} P. Feola, Xisco Jiménez Forteza, S. Capozziello, R. Cianci, S. Vignolo. \emph{Mass-radius relation for neutron stars in $f(R)=R+\alpha R^2$ gravity: A comparison between purely metric and torsion formulations}. Phys. Rev. D 101 (2020)

\bibitem{fR5} S. Capozziello, M. De Laurentis, R. Farinelli, and S. D. Odintsov, \emph{Mass-radius relation for neutron stars in f(R) gravity}. Phys. Rev. D 93 (2016) [arXiv:1509.04163]

\bibitem{fR6} D. D. Doneva, S. S. Yazadjiev, N. Stergioulas, and K. D. Kokkotas. \emph{Differentially rotating neutron stars in scalar-tensor theories of gravity}. Phys. Rev. D 98 (2018) 104039 [arXiv:1807.05449]

\bibitem{fR7} S. S. Yazadjiev, D. D. Doneva, K. D. Kokkotas. \emph{Tidal Love numbers of neutron stars in f(R) gravity}. Eur. Phys. J. C 78 (2018) 818 [arXiv:1803.09534]

\bibitem{fR8} S. D. Odintsov and V.K. Oikonomou. \emph{Inflationary attractors predictions for static neutron stars in the mass-gap region}. Phys. Rev. D 107 (2023) 104039 [arXiv:2305.05515]

\bibitem{fR9} S. S. Yazadjiev, D. D. Doneva, and K. D. Kokkotas.\emph{Oscillation modes of rapidly rotating neutron stars in scalar-tensor theories of gravity}. Phys. Rev. D 96 (2017) 064002 [arXiv:1705.06984]

\bibitem{fR10} D. D. Doneva, S. S. Yazadjiev, K. D. Kokkotas. \emph{I-Q relations for rapidly rotating neutron stars in f(R) gravity}. Phys. Rev. D 92 (2015) 064015 [arXiv:1507.00378]

\bibitem{fR11} S. S. Yazadjiev, D. D. Doneva, K. D. Kokkotas. \emph{Rapidly rotating neutron stars in R-squared gravity}. Phys. Rev. D 91 (2015) 084018 [arXiv:1501.04591]

\bibitem{fR12} K. V. Staykov, D. D. Doneva, S. S. Yazadjiev, K. D. Kokkotas. \emph{Slowly rotating neutron and strange stars in R2 gravity}. JCAP 10 (2014) 006 [arXiv:1407.2180].

\bibitem{fR13} S. S. Yazadjiev, D.D. Doneva, K.D. Kokkotas, K.V. Staykov. \emph{Non-perturbative and self-consistent models of neutron stars in R-squared gravity}. JCAP 06 (2014) 003 [arXiv:1402.4469]


\bibitem{Aldro} R. Aldrovandi and J. G. Pereira. \emph{ Teleparallel Gravity: An Introduction}. Heidelberg: Springer, (2013).

\bibitem{Bahamonde} S. Bahamonde et al. \emph{Teleparallel Gravity: From Theory to Cosmology}, Rep. Prog. Phys. 86 026901 (2023) [arXiv:2106.13793]

\bibitem{TT}Y. Cai et al., \emph{$f(T)$ teleparallel gravity and cosmology}. {Rept. Prog. Phys.} {\bf 79} (2016), 106901.

\bibitem{fR2} T.P. Sotiriou, V. Faraoni. \emph{$f(R)$ theories of gravity}. {Rev. Mod. Phys.} {\bf 82} (2010) 451.

\bibitem{Ferraro} R. Ferraro, F. Fiorini, \emph{Modified teleparallel gravity: Inflation without inflaton}. {Phys. Rev. D} {\bf 75}, (2007) 084031 [gr-qc/0610067].

\bibitem{Linder} E. V. Linder, \emph{Einstein's Other Gravity and the Acceleration of the Universe}. {Phys. Rev. D} {\bf 81}, (2010) 127301  [arXiv:1005.3039].

\bibitem{Ganiou} M. G. Ganiou et al., \emph{Strong magnetic field effects on neutron stars within f(T) theory of gravity}. {Eur. Phys. J. Plus} {\bf 132} (2017) 250.

\bibitem{Kpa}A. V. Kpadonou, M. J. S. Houndjo, M. E. Rodrigues, \emph{Tolman-Oppenheimer-Volkoff Equations and their implications for the structures of relativistic Stars in f(T) gravity}. {Astrophys. Space Sci.} {\bf 361} (2016) 244.

\bibitem{Pace} M. Pace, J. L. Said, \emph{A Perturbative Approach to Neutron Stars in f(T,T)-Gravity}. {Eur. Phys. J. C} {\bf 77} (2017) 283.

\bibitem{Ilijic} S. Iliji\'c, M. Sossich, \emph{Compact stars in $f(T)$ extended theory of gravity}. {Phys. Rev. D} {\bf 98} (2018) 064047.

\bibitem{Bohmer} C. G. Böhmer, A. Mussa and N. Tamanini, \emph{Existence of relativistic stars in f(T) gravity}. Class. Quantum Grav. {\bf 28} (2011) 245020.

\bibitem{Pace2} M. Pace, J. L. Said, \emph{Quark stars in $f(T,\mathcal{T})$-gravity}. Eur. Phys. J. C {\bf 77} (2017) 62.

\bibitem{FA1} H. G. M. Fortes, J. C. N. de Araujo, \emph{Solving Tolman-Oppenheimer-Volkoff equations in f(T) gravity: a novel approach}. Classical and Quantum Gravity, 39, 245017 (2022), [arXiv:2105.04473]   

\bibitem{FA2} J. C. N. de Araujo, H. G. M. Fortes, \emph{Solving Tolman-Oppenheimer-Volkoff equations in f(T) gravity: a novel approach applied to polytropic equations of state}, Brazilian Journal of Physics, 53, 75 (2023) [arXiv:2105.09118]   

\bibitem{FA3} J. C. N. de Araujo, H. G. M. Fortes, \emph{Solving Tolman-Oppenheimer-Volkoff equations in f(T) gravity: a novel approach applied to some realistic equations of state}. International Journal of Modern Physics D 31, 2250101 (2022)  [arXiv:2109.01155]

\bibitem{FA4} J. C. N. de Araujo, H. G. M. Fortes. \emph{Mass of compact stars in f(T) gravity}. Eur. Phys. J. C., 85, 376  (2023)[arXiv:2211.07418]  

\bibitem{Myr} R. Myrzakulov, \emph{Accelerating universe from F(T) gravity}. {Eur. Phys. J. C} {\bf 71} (2011) 1752 [arXiv:1006.1120].

\bibitem{Karami} K. Karami, A. Abdolmaleki, \emph{$f(T)$ modified teleparallel gravity models as an alternative for holographic and new agegraphic dark energy models}. {Res. Astron. Astrophys.} {\bf 13} (2013)  757.

\bibitem{SSS10} C. G. Böhmer, A. Mussa, N. Tamanini, \emph{Existence of relativistic stars in $f(T)$ gravity}. {Classical Quantum Gravity} {\bf 28} (2011) 245020.

\bibitem{Ilijic2} A. DeBenedictis, S. Iliji\'c, \emph{Spherically symmetric vacuum in covariant $F(T ) = T +\frac{\alpha}{2}T^2+O(T^\gamma)$ gravity theory}, Phys. Rev. D94 no. 12, (2016) 124025, arXiv:1609.07465 [gr-qc].

\bibitem{Nunes} R. C. Nunes, J. G. Coelho, J. C. N. de Araujo, \emph{Weighing massive neutron star with screening gravity: A look on PSR J0740+6620 and GW190814 secondary component}. {Eur. Phys. J. C} {\bf 80} (2020) 11115  [arXiv:2008.10395].

\bibitem{GW} R. Abbott et al., \emph{GW190814: Gravitational Waves from the Coalescence of a 23 Solar Mass Black Hole with a 2.6 Solar Mass Compact Object}. {The Astrophysical Journal} {\bf 896}  (2020) L44  [arXiv:2006.12611].

\bibitem{Dadhich} N. Dadhich. \emph{Buchdahl compactness limit and gravitational field energy}. Journal of Cosmology and Astroparticle Physics, 2020(04), 035–035. https://doi.org/10.1088/1475-7516/2020/04/035

\bibitem{buch1959} H. A. Buchdahl, \emph{General relativistic fluid spheres}. Physical Review, v. 116, n. 4, p. 1027–1034, (1959).

\bibitem{TOV} C. M. Will, \emph{Theory and experiment in gravitational physics}. Cambridge University Press (1993).

\bibitem{olmo} G. J. Olmo, D. Rubiera-Garcia, A. Wojnar, \emph{Stellar structure models in modified theories of gravity: Lessons and challenges}. Phys. Report {\bf 876} (2020) 1.

\bibitem{Ilijic20} S. Iliji\'c, M. Sossich, \emph{Boson stars in $f(T)$ extended theory of gravity}. {Phys. Rev. D} {\bf 102} (2020) 084019.

\bibitem{Nair} K. K. Nair, M. T. Arun, \emph{Skyrmion in teleparallel gravity}, (2023) arXiv:2307.11933 [gr-qc].

\bibitem{Bengo} G. R. Bengochea, R. Ferraro, \emph{Dark torsion as the cosmic speed-up}. {Phys. Rev. D} {\bf 79}, 124019 (2009)
[arXiv:0812.1205 [astro-ph]].

\bibitem{SSS11} R. Ferraro, F. Fiorini, \emph{Spherically symmetric static spacetimes in vacuum $f(T)$ gravity}. {Phys. Rev. D} {\bf 84} (2011) 083518.

\bibitem{SSS12} L. Iorio, E. N. Saridakis, \emph{Solar system constraints on $f(T)$ gravity}. {Mon. Not. Roy. Astron. Soc.} {\bf 427} (2012) 1555.

\bibitem{Tamanini} N. Tamanini and C. G. Böhmer, \emph{Good and bad tetrads in f(T) gravity}. {Phys. Rev. D} {\bf 86} (2012)  044009.

\bibitem{Krssak} M. Kr\v s\v s\'ak, E.N. Saridakis, \emph{The covariant formulation of f(T) gravity}. {Classical and Quantum Gravity} {\bf  33} (2016) 115009.

\bibitem{eos0} L. Ferrari, P. C. R. Rossi, M. Malheiro. \emph{A polytropic approach to neutron stars}. {International Journal of Modern Physics D} {\bf 19} (2010) 1569.

\bibitem{BS2010} T.W. Baumgarte, S.L. Shapiro. {Numerical Relativity: Solving Einstein's Equations on the Compute}, Cambridge University Press (2010).











































\end{thebibliography}
\end{document}